\documentclass[lettersize,journal]{IEEEtran}
\usepackage{booktabs}
\usepackage{diagbox}
\usepackage{makecell}
\usepackage{array}
\usepackage{graphicx,amssymb,amsmath}
\usepackage{multicol}
\usepackage[noadjust]{cite} 
\usepackage[justification=centering]{caption}
\usepackage[font=small]{caption}
\usepackage{setspace}
\usepackage{subfigure}
\usepackage{graphicx}
\usepackage{float}
\usepackage {url}
\usepackage{stfloats}
\usepackage[normal]{threeparttable}
\usepackage{amsthm,pifont}
\usepackage{flushend}
\usepackage{cases,subeqnarray}
\usepackage{bm,multirow,bigstrut}
\usepackage{amsmath, amsthm, amssymb}
\usepackage{textcomp}
\usepackage{latexsym,bm}
\usepackage{booktabs}
\usepackage{mathtools}
\usepackage{dsfont}
\usepackage{extarrows}
\usepackage{epsfig}
\usepackage{epsfig}
\usepackage{epstopdf}
\usepackage[table]{xcolor}
\usepackage{colortbl} 
\usepackage{multirow, hhline}
\usepackage{xcolor}
\usepackage{array} 
\usepackage{algorithmicx,algorithm}
\usepackage{hyperref}
\usepackage{tikz}
\usepackage{makecell}
\usepackage{enumitem}
\usepackage{comment}
\usepackage{pifont}
\usepackage{graphicx}
\usepackage{tabularray}

\usepackage{subcaption}  

\usetikzlibrary{shapes.geometric, arrows, positioning}
    
\theoremstyle{plain}

\theoremstyle{plain}

\usepackage{amsmath}

\usepackage[edges]{forest}
\usetikzlibrary{shadows}

\usetikzlibrary{shapes.geometric, arrows}

\definecolor{lightblue}{rgb}{0.68, 0.85, 0.90}
\definecolor{lightgreen}{rgb}{0.56, 0.93, 0.56}
\definecolor{lightpurple}{rgb}{0.75, 0.58, 0.92}

\tikzstyle{box} = [rectangle, rounded corners, minimum width=3cm, minimum height=1cm,text centered, draw=black, fill=lightblue]
\tikzstyle{boxgreen} = [rectangle, rounded corners, minimum width=3cm, minimum height=1cm,text centered, draw=black, fill=lightgreen]
\tikzstyle{boxpurple} = [rectangle, rounded corners, minimum width=3cm, minimum height=1cm,text centered, draw=black, fill=lightpurple]
\tikzstyle{line} = [draw, -latex]

\IEEEoverridecommandlockouts
\begin{document}
\title{ Enhancing Wireless Networks for IoT with Large Vision Models: Foundations and Applications}

\author{Yunting Xu, Jiacheng Wang, Ruichen Zhang, Dusit Niyato~\IEEEmembership{Fellow,~IEEE}, Deepu Rajan, Liang Yu, \\ Haibo~Zhou,
Abbas Jamalipour,~\IEEEmembership{Fellow,~IEEE}, and Xianbin Wang,~\IEEEmembership{Fellow,~IEEE}
\thanks{Y. Xu, J. Wang, R. Zhang, D. Niyato, and D. Rajan are with the College of Computing and Data Science, Nanyang Technological University, Singapore, 639798 (e-mail: yunting.xu@ntu.edu.sg, jiacheng.wang@ntu.edu.sg, ruichen.zhang@ntu.edu.sg, dniyato@ntu.edu.sg, asdrajan@ntu.edu.sg).}
\thanks{L. Yu is with Alibaba Cloud, China (e-mail: liangyu.yl@alibaba-inc.com).} 
\thanks{H. Zhou is with the School of Electronic Science and Engineering, Nanjing University, Nanjing, China, 210023 (e-mail: haibozhou@nju.edu.cn).}
\thanks{A. Jamalipour is with The University of Sydney, Sydney NSW 2006, Australia (e-mail: a.jamalipour@ieee.org).}
\thanks{X. Wang is with the Department of Electrical and Computer Engineering, Western University, Canada (e-mail: xianbin.wang@uwo.ca).}
}

\maketitle
\begin{abstract} Large vision models (LVMs) have emerged as a foundational paradigm in visual intelligence, achieving state-of-the-art performance across diverse visual tasks. Recent advances in LVMs have facilitated their integration into Internet of Things (IoT) scenarios, offering superior generalization and adaptability for vision-assisted network optimization. In this paper, we first investigate the functionalities and core architectures of LVMs, highlighting their capabilities across classification, segmentation, generation, and multimodal visual processing. We then explore a variety of LVM applications in wireless communications, covering representative tasks across the physical layer, network layer, and application layer. Furthermore, given the substantial model size of LVMs and the challenges of model retraining in wireless domains, we propose a progressive fine-tuning framework that incrementally adapts pretrained LVMs for joint optimization of multiple IoT tasks. A case study in low-altitude economy networks (LAENets) demonstrates the effectiveness of the proposed framework over conventional CNNs in joint beamforming and positioning tasks for Internet of drones, underscoring a promising direction for integrating LVMs into intelligent wireless systems.

\end{abstract}

\IEEEpeerreviewmaketitle

\section{Introduction}

In recent years, the integration of foundation models into visual tasks for Internet of Things (IoT) scenarios has emerged as a prevailing paradigm, exemplified by the deployment of large vision models (LVMs) such as the Vision Transformer (ViT) \cite{dosovitskiy2020image}. Different from large language models (LLMs) that primarily focus on sequential textual data, LVMs are specifically tailored for spatial and structural comprehension of visual inputs, enabling state-of-the-art performance across IoT tasks such as image recognition, semantic segmentation, and vision-language reasoning. Owing to the massive number of parameters and extensive pretraining on large-scale visual datasets, LVMs consistently outperform traditional convolutional neural networks (CNNs) in representation learning and model generalization.
Furthermore, their substantial scalability and cross-domain adaptability facilitate effective knowledge transfer across a wide range of downstream applications \cite{xu2023federated}.



Among these emerging application domains, LVM-based wireless communications have attracted considerable attention, offering promising opportunities to improve communication performance in complex and dynamic environments \cite{zhang2025embodied}.
Leveraging their advanced visual understanding capabilities, LVMs extract rich contextual information from visual inputs, enabling a range of vision-assisted communication tasks such as channel estimation, beamforming, near-field signal processing, and semantic-aware resource allocation \cite{charan2023millimeter}. The vision-derived information offers a complementary alternative to conventional radio frequency measurements, thereby significantly enhancing the system's perception of complex wireless environments.
Moreover, LVMs demonstrate robust generalization and adaptability 
for the simultaneous optimization of multiple IoT tasks. The capability underscores the potential of LVMs to serve as a unified foundational backbone for vision-assisted, multi-task wireless communication systems.

Despite the promising performance of LVMs in IoT task optimization, several implementation challenges still remain in realizing their full potential for wireless deployment scenarios. On the one hand, the large model size of LVMs leads to significant computational overhead and energy consumption during the training process, making full retraining from scratch both resource-intensive and challenging to converge. On the other hand, IoT tasks in wireless communication environments often suffer from a lack of task-specific and large-scale labeled datasets, which limits the ability to effectively adapt LVMs to domain-specific scenarios. These constraints hinder the direct deployment of LVMs in wireless networks and necessitate efficient adaptation strategies tailored to data-scarce and resource-constrained environments.


Motivated by the increasing imperative to exploit the full potential of LVMs in wireless IoT task optimization, this paper contributes to the field through investigating the architectures and functionalities of LVMs.
To address the practical challenges of adapting LVMs to wireless networks, we redesign the model’s output structure to facilitate joint optimization across multiple wireless communication tasks.
Furthermore, we propose a progressive fine-tuning framework that incrementally unfreezes the LVM layers during the training process.
The main contributions of this paper are summarized as follows.

\begin{itemize}
    \item We investigate state-of-the-art LVM architectures and analyze their functional characteristics in the context of visual understanding tasks, highlighting their generalization and adaptability across diverse vision applications.
    \item We systematically explore the integration of LVMs into wireless networks, with emphasis on their utility across the physical layer, network layer, vision-based semantic communications, and user-centric service applications. 
    \item We propose a fine-tuning framework for LVMs based on a multi-stage unfreezing pipeline, designed to support wireless multi-task optimization. A case study conducted in low-altitude economy networks (LAENets) demonstrates the performance advantages of the proposed framework over traditional CNN approaches for Internet of drones.
\end{itemize}



\section{Foundations of large vision model}



This section examines the architectural and functional properties of LVMs. Representative designs and application domains are visualized in Figure \ref{Section2_LVM}, and a detailed overview of state-of-the-art LVMs is provided in Table \ref{table1}.




\subsection{Functionality of Large Vision Models}





\subsubsection{ Image Classification}

Image classification aims to predict the category of input images. It directly processes the entire image and outputs a single class label, identifying the primary object or scene type. Unlike conventional CNN-based approaches, LVMs leverage transformer-based architectures, such as ViT and Swin Transformer, which capture global spatial patterns through patch-based self-attention mechanisms. 
These models offer superior accuracy and robustness when handling diverse and complex image datasets.



\subsubsection{ Object Detection} 


Object detection involves identifying and localizing multiple discrete entities within an image. Detection models output precise spatial coordinates, represented as bounding boxes, along with semantic labels and associated confidence scores. Recent models such as DETR leverage an encoder-decoder-based transformer architecture to enable efficient multi-object detection. 
This functionality enhances both spatial and semantic understanding, facilitating applications such as autonomous navigation and real-time monitoring.





\subsubsection{ Image Segmentation}
Image segmentation partitions an image into coherent, semantically meaningful regions at a pixel-level resolution. Recent LVMs, exemplified by Segment
Anything Model (SAM), employ prompt-based, category-agnostic methods to effectively segment arbitrary and unseen objects. The generalization capability of LVM significantly extends segmentation versatility and applicability across diverse and previously unencountered scenarios.


\subsubsection{ Image reconstruction}

Image reconstruction tasks focus on recovering missing, corrupted, or partially occluded image content. Leveraging self-supervised learning approaches, advanced LVMs such as Masked Autoencoder (MAE) employ Autoencoder to reconstruct occluded image regions through sophisticated masked encoding and decoding frameworks. 
The self-supervised pretraining paradigm of LVM enhances representation robustness, enabling effective learning from limited or incomplete data, even in the absence of labels.


\subsubsection{ Image representation}


Visual representation learning seeks to extract semantically rich, transferable feature embeddings from images. 
LVMs such as DINO employ self-distillation techniques to efficiently generate meaningful visual representations without the need for explicit annotations.
The representation capability of LVM can significantly enhance downstream performance across various vision-assisted tasks.



\subsubsection{Generation Capability}


Generative LVMs synthesize novel content based on learned patterns and data distributions. Applications include BLIP and Diffusion Transformer (DiT) that generates natural language descriptions or diverse image variants under varied conditions, such as changes in weather or lighting. This generation capability of LVMs facilitates data augmentation and enhances model robustness. 

\subsubsection{Multimodal and multi-task learning}


Multimodal learning like CLIP integrates multiple data modalities, such as vision and language, to enable sophisticated cross-modal reasoning in a joint embedding space. 
Besides, unified frameworks like One-For-All (OFA) simultaneously handle diverse tasks, including classification, captioning, generation, within a single cohesive architecture, thus eliminating the need for deploying different task-specific models.



\begin{figure*}
\centering
\includegraphics [width=\textwidth]{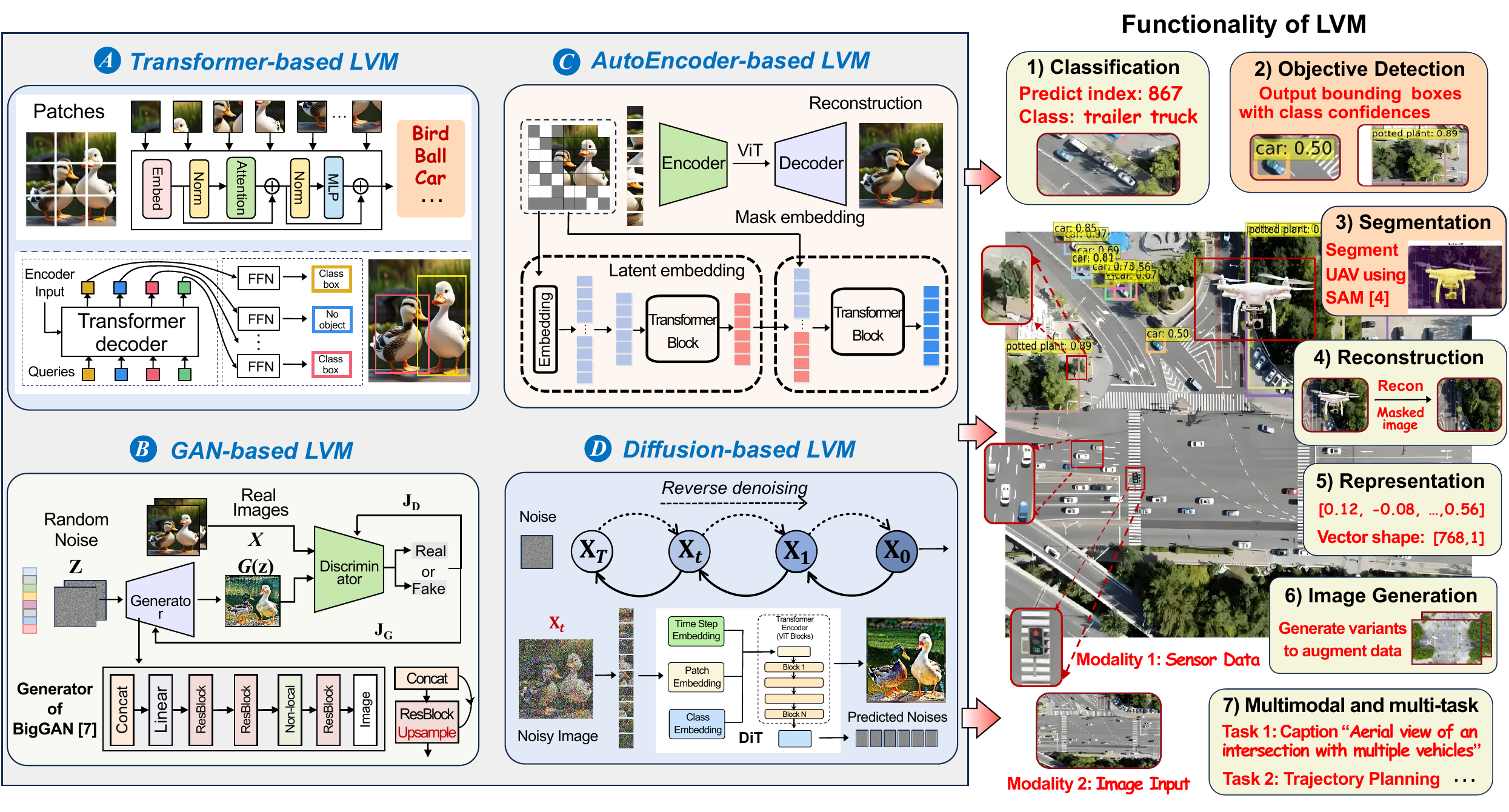} 
\captionsetup{justification=justified,format=plain}
\caption{ Representative architectures and functionalities of state-of-the-art Large Vision Models (LVMs). Left: Four major LVM paradigms, including Transformer-based, GAN-based, AutoEncoder-based, and Diffusion-based architectures. Right: Key capabilities of LVM, including classification, object detection, segmentation, reconstruction, semantic representation, image generation, multimodal and multi-task learning.
}
\label{Section2_LVM}
\end{figure*}

\subsection{Neural Network Foundations of Large Vision Models}
The significant capabilities of LVMs are attributed to the foundational neural network architectures such as Transformers, Generative Adversarial Networks (GANs), Autoencoders (AEs), and Diffusion Models (DMs), each offering unique benefits crucial to advanced visual processing tasks.


\subsubsection{\textbf{Transformer}}
Transformer architectures, originally developed for natural language processing, have become the dominant paradigm in vision modeling. As illustrated in Figure \ref{Section2_LVM}, models including ViT and Swin Transformer adopt encoder-based designs that process images as sequences of patches, while DETR introduces an encoder-decoder framework for end-to-end object detection. These transformer-based LVMs, employing self-attention mechanisms to model intricate global dependencies and hierarchically structured visual information, have consistently outperformed traditional CNNs in tasks such as image classification and object detection.

\subsubsection{\textbf{Generative Adversarial Networks}} 




GANs employ adversarial training to generate highly realistic synthetic images, which consist of a generator that synthesizes realistic data samples and a discriminator that differentiates between real and synthetic inputs. Representative GAN-based LVMs include StyleGAN\footnote{https://github.com/NVlabs/stylegan}, known for style-based photorealistic image synthesis, and BigGAN\footnote{https://github.com/ajbrock/BigGAN-PyTorch}, recognized for its high-quality large-scale image generation, both significantly impacting generative visual tasks. 
GAN-based LVMs have been widely applied in data augmentation, style transfer, and image editing.


\subsubsection{\textbf{Autoencoders}}

AEs are neural networks designed to learn compact visual representations by encoding input data into a latent space and subsequently reconstructing images through a decoding process. The encoder-decoder framework enables the model to capture essential features while discarding redundant information, which is highly effective for self-supervised learning tasks. LVMs have extensively employed AEs to implement masking and reconstruction strategies, where parts of the input image are intentionally hidden during training, forcing the model to infer missing regions from visible context. Such models are instrumental in scenarios with limited labeled data, providing strong capabilities for image reconstruction and representation learning.





\subsubsection{\textbf{Diffusion Models}}

DMs consist of a forward process, where random noise is progressively added, and a reverse process that reconstructs structured data through iterative denoising. 
A typical architecture for DM-based LVMs implements denoising processes with transformer encoders, leveraging self-attention mechanisms to predict noise across all image patches simultaneously.
Notable examples, such as DiT, have achieved state-of-the-art performance in image synthesis and restoration. DM-based LVMs have gained prominence in generative tasks,
particularly in domains such as text-to-image synthesis and high-fidelity image restoration.



\begin{table*}[t]
    \centering
    \caption{ Overview of Representative Large Vision Models. }
    \label{table1}	
    \includegraphics[width=1\textwidth]{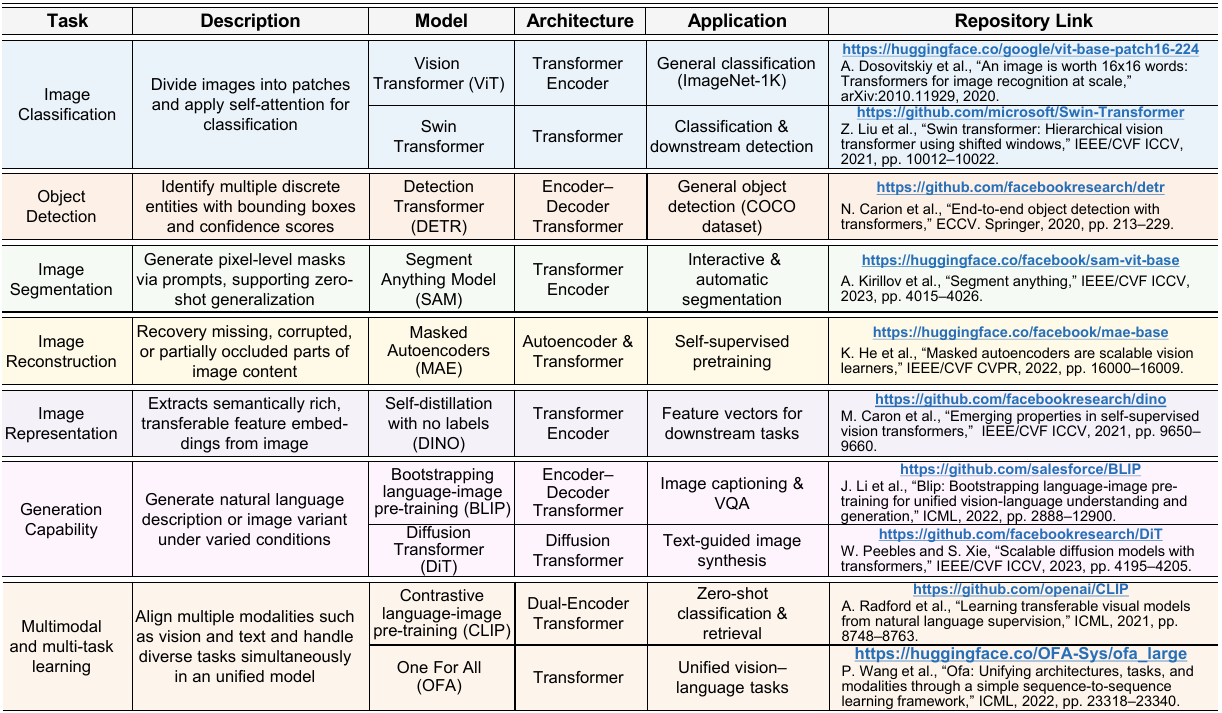} 
\end{table*}


\subsection{Key Insights of LVMs}
Based on the foundations of the LVMs, the following key insights can be summarized.


\begin{itemize}
    \item
\textit{Significant Performance Gain:} Compared with traditional vision models, LVMs exhibit enhanced generalization abilities due to their deeper architectures. For instance, ViT-B/16 \cite{chen2021vision} achieves a 5.3\% improvement in ImageNet-1K Top-1 accuracy over comparable ResNet models.
    \item \textit{Integrated Multi-Task Capability:} Unified architectures of LVMs facilitate joint training across classification, detection, segmentation, and generation within a single model, eliminating reliance on disparate specialist networks and streamlining deployment in multi-functional systems.

    \item \textit{Cross-Domain Application:} Beyond conventional vision tasks, LVMs exhibit robust adaptability to cross-domains, such as wireless communication,
    medical imaging, and biological analysis,
    underscoring their potential as universal pattern learners across heterogeneous data types.
    

\end{itemize}

\section{Applications of LVM in Wireless Networks}



This section explores the applications of LVMs in wireless networks, covering the physical layer, network layer, semantic communications and user-centric services.




\subsection{Physical Layer Communications}

\subsubsection{\textbf{Channel Modeling and Prediction} } 
    LVMs can utilize wireless environmental images to identify potential scatterers, such as buildings, vehicles, and obstacles. By mapping these visual features to radio propagation models, LVMs estimate parameters like path loss exponent, delay spread, and received power. For instance, the authors in \cite{zhang2025vision} employed YOLOv8-based object detection to identify vehicles in RGB images, enabling precise channel prediction purely from visual input and reducing reliance on traditional RF measurements.


\subsubsection{\textit{\textbf{Beamforming Selection}} }
    Beamforming and directional transmission are critical for mmWave and massive MIMO systems. 
    Traditional beam alignment requires searching through many beam directions, incurring high training overhead. 
    LVMs can use image feeds to detect blockage and identify the line-of-sight (LoS) path, predicting the optimal beam directions. Leveraging vision models on camera images for mmWave beamforming has been shown to reduce the beam search time by 93\% compared to exhaustive beam scanning method \cite{salehi2020machine}.
    

\subsubsection{\textit{\textbf{ Wireless Signal Processing}} } 
    Raw RF signals can be transformed into visual formats that are compatible with LVM processing. For instance, In-phase and Quadrature (I/Q) data streams are transformed into time-frequency spectrograms via
    constellation diagrams that visualize complex signal points in an I-Q plane \cite{gao2023moe}.
    In more advanced applications, such as near-field communications and holographic MIMO systems, signals are represented as spatial distributions of electromagnetic fields, typically formatted as 2D or 3D grids. These volumetric representations enable LVMs to directly process spatial field patterns, supporting tasks such as modulation recognition, interference detection, and signal classification.


\subsection{Network Layer}

Beyond physical layer link enhancements, LVMs can aligning network resources with spatial image patterns, facilitate routing configurations, and identify network anomalies to support adaptive detection of network attacks or failures.



\subsubsection{ \textit{\textbf{Wireless Resource Allocation}} }
    Visual images can be used to assess user density and movement patterns, enabling efficient optimization of wireless resources such as bandwidth, spectrum, and transmit power. For example, through identifying
    potential congestion via aerial or street-level imagery, LVMs can locate active users and schedule frequency and time slots to match user demand \cite{tian2020applying}.
    Additionally, by tracking user motion information such as position and speed, camera-based methods can proactively predict optimization strategies, enabling resource reservation and timely allocation.

\subsubsection{ \textit{\textbf{Routing and Topology Planning}} }
    In multi-hop networks such as vehicular ad-hoc network, vision extracts topology information through environmental cues such as congested roads and identifiable landmarks, allowing the network to route data via vehicle nodes along unobstructed paths.
    By utilizing visual topology information to design advanced routing algorithms, rather than relying solely on shortest-hop routing, the network can significantly reduce end-to-end delay, minimize packet loss, and enhance overall throughput.
        
\subsubsection{ \textit{\textbf{Network Security}} }
    Network flows and telemetry can be visualized in diverse forms, including time-series of traffic load and heat maps of network usage. Leveraging the recognition capabilities of LVMs, abnormal patterns such as DDoS attacks and traffic anomalies can be identified from visual spatio-temporal representations. Unlike threshold-based monitoring, LVMs can automatically detect abnormal features
    and provide rapid response to potential outages or security threats. Vision-aided anomaly detection has shown to achieve over 99.88\% traffic classification accuracy  while improving throughput by 10\% compared to traditional threshold methods \cite{liu2023real}.





\subsection{Vision-Based Semantic Communications}

Semantic communication transmits representative messages rather than raw bits. Leveraging LVMs to compress and encode visual content allows only the most essential information to be transmitted, which further improves the efficiency of semantic communication under strict bandwidth and latency constraints.

\begin{figure*}
\centering
\includegraphics [width=\textwidth]{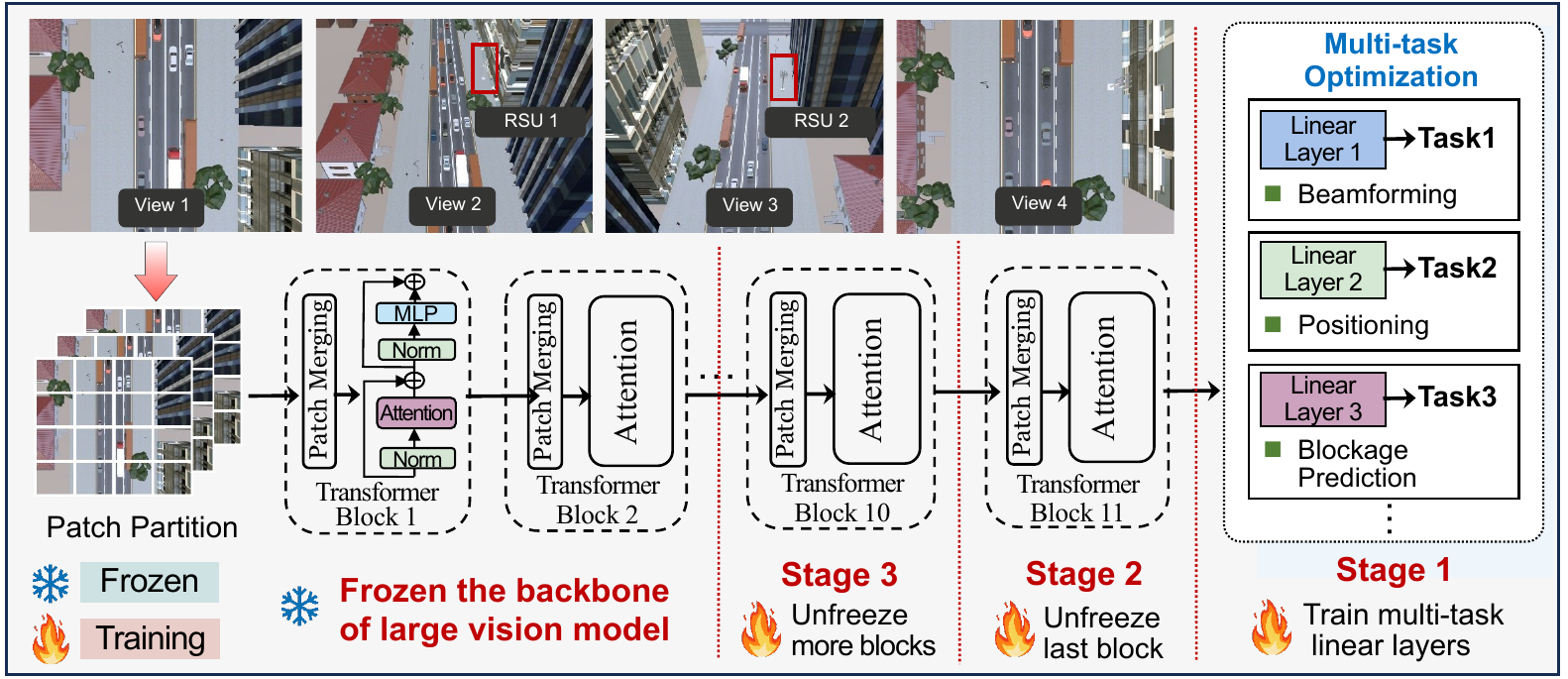} 
\captionsetup{justification=justified,format=plain}
\caption{ The proposed fine-tuning LVM framework for multi-task network optimization. The training process follows a multi-stage pipeline: Stage 1 trains only the task-specific linear heads with the backbone frozen; Stage 2 unfreezes the last Transformer block for joint optimization; and Stage 3 progressively unfreezes additional blocks for full fine-tuning. The red rectangles highlight the RSUs in View 2 and View 3.
}
\label{Section4_framework}
\end{figure*}

\subsubsection{ \textit{\textbf{Semantic Feature Extraction and Transmission}} }
Instead of transmitting full-resolution images, LVMs distill essential semantics, such as object identities, locations, and contextual information that are critical for downstream tasks. For example, the authors in \cite{park2024vision} present a ViT-based framework with importance-aware quantization, which quantifies the significance of image patches and assigns more transmission bits to patches with high attention scores. Compared to conventional fixed-rate quantization schemes, LVM-based method achieves up to 38.5\% improvement in image classification accuracy and significantly reduce the communication overhead.


\subsubsection{ \textit{\textbf{Joint Source-Channel Coding (JSCC)}} }
    LVMs enhance semantic communication in channel coding by directly mapping camera images into compact latent features for transmission over noisy channels. Unlike traditional vision models that separate source and channel coding, LVM-based JSCC jointly optimizes semantic compression and error resilience in an end-to-end manner. For example, the authors in \cite{10589474} employ Swin Transformer as the encoder at the transmitter to produce semantically rich latent codes. The receiver-side decoder reconstructs the image while compensating for channel noise. Through dynamically adjusting to channel conditions, a 1.3 dB peak SNR gain is achieved versus CNN-based baselines. 

\subsubsection{ \textit{\textbf{ Multimodal LVM-Assisted Semantic Communication}} }
    A vision-language model can generate concise captions or descriptions that summarize the key content of an image. 
    Except for transmitting extracted visual features, multimodal LVMs combine textual summary over the wireless channel, allowing the receiver to reconstruct the content using both vision and language cues via a single model. 
    This strategy integrates multiple modalities that together convey the image semantics in a highly compressed form, enhancing bandwidth efficiency and robustness under varied wireless conditions.
    


\subsection{User-centric Service Applications}
LVMs enhance user-centric applications by integrating visual intelligence with network data, enabling high-precision localization, user trajectory optimization and hotspot prediction, as well as supporting the creation of wireless digital twins.

\subsubsection{ \textit{\textbf{High-Precision Localization}}}
    LVMs can fuse visual cues with wireless signals to achieve centimeter-level localization, enabling applications such as indoor navigation, autonomous driving, and industrial automation.
    For example, vision encoders like DINOv2 extract robust spatial features from complex environments, including urban streets or industrial sites, enhancing localization accuracy even under challenging conditions such as occlusion or poor lighting \cite{yang2025dinov2}. 
    
\subsubsection{ \textit{\textbf{Trajectory and Hotspot Prediction}} } LVMs are capable of predicting user trajectories and usage hotspots by tracking objects and motion in camera feeds. Vision-based models learn patterns of crowd movement and pedestrian flow, enabling anticipatory load balancing, such as steering traffic to underused APs and hotspot-aware content caching. Applications like large events or transit hubs are facilitated by LVMs, where early detection of congestion can significantly enhance operational efficiency.

    
\subsubsection{ \textit{\textbf{Wireless Digital Twin Creation}} }
    LVMs can process large-scale visual data to build detailed 3D models of real-world environments, forming the foundation of wireless digital twins. These models capture structural layouts, surface materials, and spatial geometry, which are critical for simulating signal propagation and interference. By combining visual data with wireless measurements, LVM-assisted digital twins enable accurate prediction of network performance in complex settings. This approach allows network planners to virtually test and optimize deployments, reducing reliance on costly and time-consuming physical trials.

\section{Proposed Fine-tuning LVM Framework for Multi-task Network Optimization}


To leverage the advanced capabilities of LVMs in wireless communication scenarios, we propose a tailored fine-tuning framework as presented in Figure \ref{Section4_framework} for multi-task network optimization. 
While traditional CNN-based methods excel in specific, single-wireless task, they typically face challenges in multi-task optimization, primarily due to the constrained representational capacity and limited adaptability across varying scenarios.
In contrast, LVMs, with their intrinsic versatility and generalized ability, are capable of effectively handling diverse wireless tasks simultaneously, offering a promising solution for multi-task wireless network optimization.

Nevertheless, directly training LVMs from scratch for wireless communication tasks poses practical difficulties. 
The primary challenges arise from the large model size, which complicates model training convergence and significantly increases computational complexity. Additionally, datasets specific to wireless tasks are typically limited in size and diversity, making it challenging to adequately train  LVMs with a substantial number of parameters.
To address these limitations, we perform a progressive fine-tuning strategy to pretrained LVMs, facilitating efficient adaptation to multi-task wireless network optimization. As illustrated in Figure \ref{Section4_framework}, the training process of LVMs follows a multi-stage fine-tuning pipeline. Specifically, 

\begin{itemize}
    \item Stage 1 (Task-specific linear head training): The training parameters restricted solely to task-specific linear heads, with the LVM backbone entirely frozen, which preserves pretrained representations while enabling fast adaptation to wireless tasks with minimal computational overhead.

    \item Stage 2 (Selective backbone fine-tuning): The final block of an LVM is unfrozen and jointly optimized  with task-specific linear heads, promoting targeted representation refinement and enhancing task-specific adaptability. 

    \item Stage 3 (Incremental comprehensive fine-tuning): Additional LVM blocks are progressively unfrozen to enable deeper, comprehensive fine-tuning. This stepwise procedure systematically optimizes model parameters, significantly aligning learned representations with complex, multi-task wireless network optimization objectives.
\end{itemize}

In the proposed framework, the output layers of LVMs is redesigned with multiple linear heads to simultaneously support diverse wireless tasks, including beamforming, positioning, and blockage prediction.
Task-specific loss functions need to be modeled, such as regression for positioning, multi-class prediction for beamforming, and binary classification for blockage detection. These loss functions would further be combined with weighting schemes to effectively realize the potential of pretrained LVMs within the fine-tuning framework. 

\section{Case Study}

\begin{figure}
\centering
\includegraphics [width=0.495\textwidth]{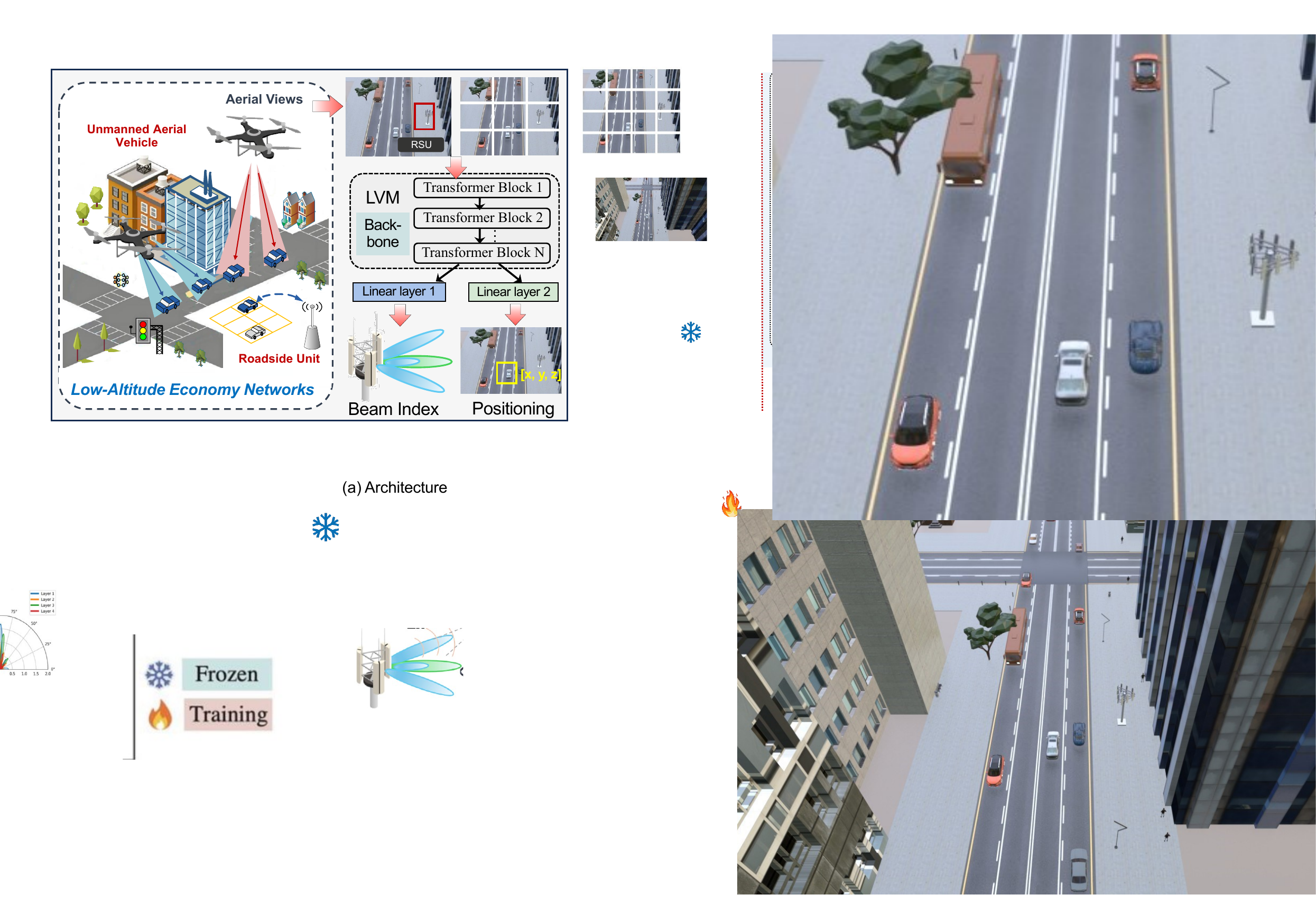} 
\captionsetup{justification=justified,format=plain}
\caption{ Joint optimization of beamforming selection and user positioning for Internet of drones under LAENet scenario.
}
\label{Section5_case_scenario}
\end{figure}

In this section, to validate the effectiveness of the proposed fine-tuning LVM framework, we conduct a case study focused on joint beamforming selection and user positioning for Internet of drones under LAENet scenario as illustrated in Figure~\ref{Section5_case_scenario}.

\subsection{Scenario Description}
We consider an urban street scenario consisting of vehicular users and roadside units (RSUs). Aerial images captured by drones are processed by an LVM to simultaneously perform two tasks: (i) predicting the optimal beam index based on the Type-I codebook defined in 3GPP 38.214 \cite{xu2025fully} for transmission, and (ii) estimating the user's position for localization. To evaluate the effectiveness of the predicted precoding decisions, we compute the achievable rate after channel equalization for vehicular users. In this setup, the RSU is equipped with an 8$\times$2 dual-polarized uniform planar array (UPA), while the vehicle is equipped with a single-antenna uniform linear array (ULA). For localization accuracy, we use the Euclidean distance between the predicted and ground-truth positions to calculate the average positioning error. 

\subsection{Dataset and Baseline Methods}

We conduct experiments on the publicly available ViWi-Drone dataset \cite{alrabeiah2020viwi}. The ViWi-Drone dataset is a large-scale, vision-aided wireless communication dataset, which contains 6,735 data samples captured by drone for model training. Each sample includes aerial-view RGB images, 3D spatial positioning information, and corresponding beamforming labels based on the 3GPP-defined codebook. These aerial-view images are collected at an altitude of 50 meters based on 17 distinct drone trajectories. For the simulation setup, we split the data samples into 70\% for training and 30\% for validation.

To assess the performance of the fine-tuning framework, we compare traditional vision model and several LVM backbones:
\begin{itemize}
    \item \textbf{ResNet-50}\footnote{https://huggingface.co/microsoft/resnet-50} is a traditional CNN model. Due to its simpl architecture and fast converge characteristic, ResNet-50 is trained from scratch using the task-specific data. 
    \item \textbf{ViT-B/16}\footnote{https://huggingface.co/google/vit-base-patch16-224}, \textbf{Swin Transformer}\footnote{https://github.com/microsoft/Swin-Transformer}, and \textbf{DINOv2}\footnote{https://github.com/facebookresearch/dinov2} are LVM backbones. We perform a three-stage fine-tuning strategy on these LVMs to efficiently adapt the pretrained models to domain-specific wireless tasks, including (i) Train the task-specific linear prediction heads; (ii) Unfrozen the final Transformer block for joint optimization; (iii) Unfrozen the penultimate block for further adaptation.
\end{itemize}

Both ResNet-50 and LVMs are evaluated under single-task and multi-task settings. In the single-task setting, these models are trained for beamforming prediction, where supervised learning is used to output the optimal beam index based on visual input. In the multi-task setting, beamforming prediction and user localization tasks are jointly optimized.
We formulate the beamforming prediction task as a classification problem with a cross-entropy loss, while the localization task is considered to be a regression problem using a mean squared error (MSE) loss. The total loss is computed as a weighted sum of the two individual loss functions.

\begin{figure*}[htbp]
    \centering
    \begin{minipage}{0.45\textwidth}
        \centering
        \includegraphics[width=\linewidth]{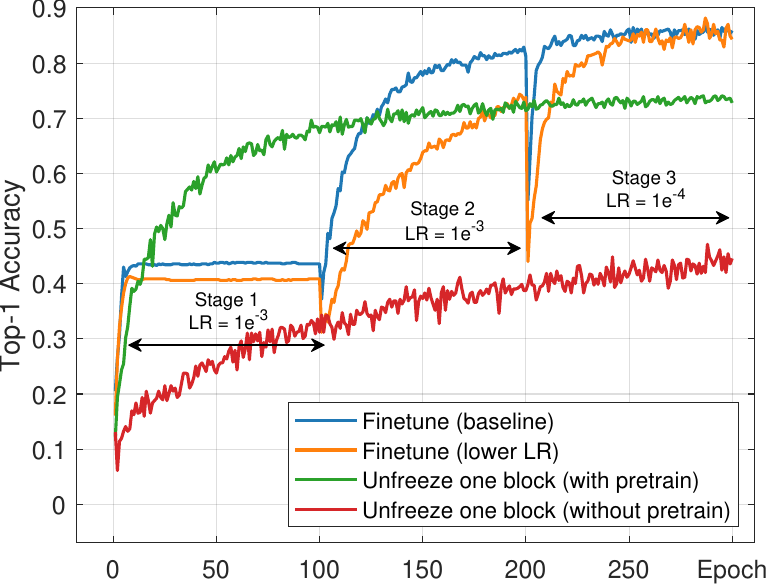}
        \captionof{figure}{ Top-1 Accuracy over epochs based on DINOv2.}
        \label{Simulation_3}
    \end{minipage}
    \hspace{0.04\textwidth} 
    \begin{minipage}{0.486\textwidth}
        \centering
        \includegraphics[width=\linewidth]{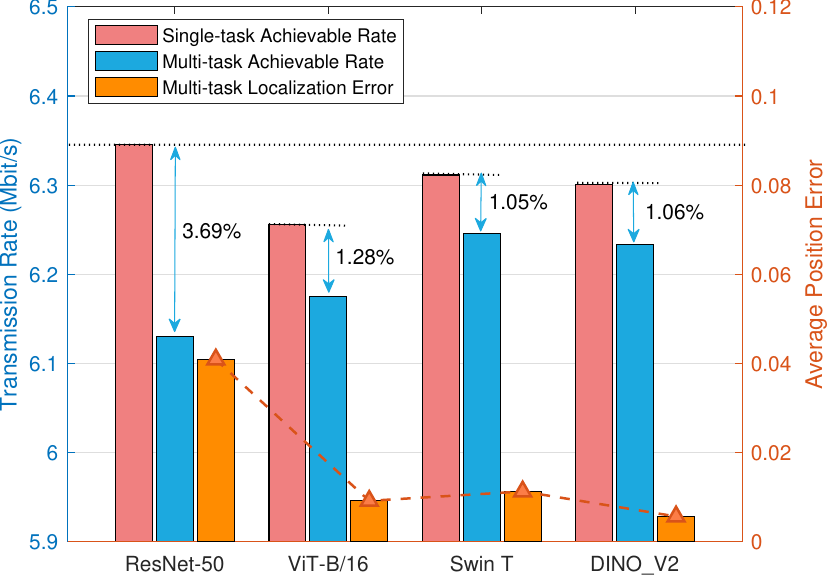}
        \captionof{figure}{Performance comparison across different LVMs.}
        \label{Simulation_4}
    \end{minipage}
\end{figure*}

\subsection{Performance Evaluation} 



Figure~\ref{Simulation_3} illustrates the training process of DINOv2 under different training and learning rate configurations.  Specifically, our proposed 
fine-tuning strategy significantly improve convergence and top-1 beam prediction accuracy for both two learning rate settings. In addition, directly fine-tuning LVMs from the beginning results in slower convergence and suboptimal performance, showcasing the effectiveness of multi-staged adaptation for pretrained LVMs in wireless tasks. Furthermore, retraining LVMs without pretrained weights leads to lower accuracy and unstable learning behavior. 
This performance limitation stems from the large number of parameters in LVMs, which poses significant  challenges when trained from scratch.


Figure~\ref{Simulation_4} compares the single-task and multi-task performance of different vision backbones in terms of achievable transmission rate and localization accuracy. The traditional CNN-based model (ResNet-50) achieves the highest rate in the single-task setting but suffers the largest performance drop (3.69\%) when extended to multi-task learning. In contrast, LVMs (ViT-B/16, Swin-Transformer, and DINOv2) demonstrate more robust performance under multi-task settings. Specifically,
multi-task learning of LVMs slightly reduces communication performance (1.05\%–1.28\%) but provides substantial benefits in localization accuracy, validating the efficiency of LVM-based joint wireless task optimization.






\section{Conclusion}
In this paper, we have analyzed the foundational architectures and functionalities of LVMs in various visual tasks, such as classification, segmentation, generation, and multimodal reasoning. Building on these insights, we have explored LVM applications across the physical layer, network layer, semantic communication, and user-centric services. To address the challenges posed by the large-scale model size and limited training data in wireless networks, we have proposed a progressive fine-tuning framework that incrementally adapts pretrained LVMs for multi-task optimization. Future work will focus on leveraging innovative AI techniques such as chain-of-thought and knowledge distillation to further facilitate LVMs as a core component in next-generation intelligent wireless systems.


\bibliographystyle{IEEEtran}
\bibliography{IEEEabrv,Ref}


\end{document}